%% file: main.tex
\documentclass[acmsmall]{acmart}
\makeatletter
\def\@ACM@checkaffil{
    \if@ACM@instpresent\else
    \ClassWarningNoLine{\@classname}{No institution present for an affiliation}%
    \fi
    \if@ACM@citypresent\else
    \ClassWarningNoLine{\@classname}{No city present for an affiliation}%
    \fi
    \if@ACM@countrypresent\else
        \ClassWarningNoLine{\@classname}{No country present for an affiliation}%
    \fi
}
\makeatother

\usepackage{caption}
\usepackage{subcaption}
\usepackage{tikz}
\usepackage{pgf-pie}
\usepackage{xcolor}
\usepackage{enumitem}
\usepackage{tcolorbox}
\usepackage{listings}
\usepackage{xspace}
\usepackage{wrapfig}
\usepackage{fancyvrb}
\usepackage[colorinlistoftodos,prependcaption,textsize=tiny]{todonotes}

\AtBeginDocument{%
  }

\lstdefinelanguage{diff}{
    morecomment=[f][\color{gray}]{@@},    
    morecomment=[f][\color{black}]{---},  
    morecomment=[f][\color{black}]{+++},  
    morecomment=[f][\color{magenta}]{**}, 
    morecomment=[l][\color{blue}]{+},     
    morecomment=[l][\color{red}]{-},      
}

\lstdefinestyle{diffstyle}{
    language=diff,
    basicstyle=\small,
    keywordstyle=\color{blue}\bfseries,
    commentstyle=\color{gray},
    numbers=left,
    numberstyle=\tiny\color{gray},
    stepnumber=1,
    numbersep=10pt,
    showstringspaces=false,
    breaklines=true,
    breakatwhitespace=false,
    captionpos=b,
    frame=single, 
}

\begin{document}

\title{Fixing Security Vulnerabilities with AI in OSS-Fuzz}

\acmArticleType{Review}


\input{macro}

\author{Yuntong Zhang}
\email{yuntong@comp.nus.edu.sg}
\affiliation{
  \institution{National University of Singapore}
}
\author{Jiawei Wang }
\email{wangjw@comp.nus.edu.sg}
\affiliation{
  \institution{National University of Singapore}
}
\author{Dominic Berzin }
\email{dominic.berzin@u.nus.edu}
\affiliation{
  \institution{ National University of Singapore}
}

\author{Martin Mirchev}
\email{mmirchev@comp.nus.edu.sg}
\affiliation{
  \institution{National University of Singapore}
}
\author{Dongge Liu}
\email{donggeliu@google.com}
\affiliation{
  \institution{Google}
}
\author{Abhishek Arya}
\email{aarya@google.com}
\affiliation{
  \institution{Google}
}
\author{Oliver Chang }
\email{ochang@google.com}
\affiliation{
  \institution{Google}
}
\author{Abhik Roychoudhury}
\email{abhik@comp.nus.edu.sg}
\affiliation{
  \institution{National University of Singapore}
}
\renewcommand{\shortauthors}{Zhang et al.}
\begin{abstract}
Critical open source software systems undergo significant validation in the form of lengthy fuzz campaigns. The fuzz campaigns typically conduct a biased random search over the domain of program inputs, to find inputs which crash the software system. Such fuzzing is useful to enhance the security of software systems in general since even closed source software may use open source components. Hence testing open source software is of paramount importance. Currently OSS-Fuzz is the most significant and widely used infrastructure for continuous validation of open source systems. Unfortunately even though OSS-Fuzz has identified more than 10,000 vulnerabilities across 1000 or more software projects, the detected vulnerabilities may remain unpatched, as vulnerability fixing is often manual in practice. 

In this work, we rely on the recent progress in Large Language Model (LLM) agents for autonomous program improvement including bug fixing. This is also the first such study with large-scale vulnerability fixing on real projects to the best of our knowledge.
We customise the well-known AutoCodeRover agent for fixing security vulnerabilities. This is because LLM agents like AutoCodeRover fix bugs from issue descriptions,  via code search. Instead for security patching, we rely on the test execution of the exploit input to extract code elements relevant to the fix. Our experience with the vulnerability data from OSS-Fuzz leads us to many observations. We note that having autonomy in the LLM agent is useful for successful security patching, as opposed to approaches like Agentless where the control flow is fixed. More importantly our findings show that we cannot measure quality of patches by code similarity of the patch with reference codes (as in CodeBLEU scores used in VulMaster), since patches with high CodeBLEU scores still fail to pass given the given exploit input. Our findings indicate that security patch correctness needs to consider dynamic attributes like test executions as opposed to relying of standard text/code similarity metrics. 
\end{abstract}

\maketitle

\input{intro}

\input{ossfuzz}

\input{methodology}

\input{evaluation}

\input{related}

\input{conclusion}

\section*{Data Availability}

Our experimental artifacts  can be found at \\
\url{https://github.com/nus-apr/code-rover-s-artifacts}.

\section*{Acknowledgments}

This work is partially supported by the Singapore Ministry of Education (MoE) Tier3 research grant "Automated Program Repair" MOE-MOET32021-0001.
This work is partially supported by the National Research Foundation, Singapore, and Cyber Security Agency of Singapore under its National Cybersecurity R\&D Programme (Fuzz Testing <NRF-NCR25-Fuzz-0001>).
Any opinions, findings and conclusions, or recommendations expressed in this material are those of the author(s) and do not reflect the views of Singapore Ministry of Education, National Research Foundation,
Singapore, and Cyber Security Agency of Singapore.

\bibliographystyle{acm}
\bibliography{reference}
\end{document}

%% file: macro.tex
\newcommand{\formattool}[1]{\textsc{#1}\xspace}

\newcommand{\tool}{\formattool{CodeRover-S}}

\newcommand{\acr}{\formattool{AutoCodeRover}}
\newcommand{\sweagent}{\formattool{SWE-Agent}}
\newcommand{\specrover}{\formattool{SpecRover}}
\newcommand{\agentless}{\formattool{Agentless}}
\newcommand{\vulmaster}{\formattool{VulMaster}}

\newcommand{\ossfuzz}{OSS-Fuzz\xspace}

%% file: intro.tex
\section{Introduction}

Security vulnerabilities are one of the major threats to modern software systems. 
Once exploited by malicious attackers, security vulnerabilities can cause significant damage to the software and its users, incurring financial loss, data breaches, and more.
In 2023, 30,927 new Common Vulnerabilities and Exposures (CVEs) are recorded by the National Vulnerability Database (NVD), and half of these vulnerabilities were classified as high or critical severity~\cite{skybox-report}.
The number of new CVEs has increased by 17\% compared to the previous year, underscoring the accelerated pace of vulnerability detection and the critical need for timely remediation.
The recent advancement in automatic programming with generative AI could further exacerbate the security issues, since some parts of the application code could come from Large Language Models (LLMs) with little security assurance.

To safeguard the software systems, researchers and practitioners have made advances in both vulnerability detection and remediation.
To detect security vulnerabilities before they are discovered/exploited by attackers, various techniques from static analysis~\cite{semgrep, codeql} to fuzzing~\cite{libfuzzer, afl} have been developed and also adopted in the industry. 
Static analysis techniques can be applied to detect a wide range of vulnerabilities. 
However, they are known to report false-positive warnings since they are often based on abstraction and conservative approximation of the program semantics~\cite{guo2023mitigating}.
Fuzzing, on the other hand, employs a biased random search in the program's input space and dynamically executes the program.
The dynamic nature of fuzzing ensures that a reported bug is a true positive.
Fuzzing has been employed by major software companies to continuously scan for vulnerabilities in their development process~\cite{chromium-security, onefuzz}.
Google's \ossfuzz, announced in 2016, provides continuous fuzzing for various core open-source software~\cite{ossfuzz-announced}.
As of August 2023, \ossfuzz has identified over 10,000 vulnerabilities across 1,000 projects~\cite{ossfuzz}.

While vulnerability detection techniques like fuzzing have shown to be both mature and effective, detection is only the first step in comprehensive software protection. 
A detected bug should be patched as soon as possible to reduce the time of exposure and the risk of being exploited.
A previous study in 2021 has shown that the median time-to-fix (i.e. time from bug reporting to patch verification) to be 5.3 days for bugs detected by \ossfuzz~\cite{ding2021empirical}, and 10\% of the reported bugs are not fixed within the 90-day disclosure deadline.
The rising number of detected vulnerabilities in recent years may require developers to invest even more time and effort in manually patching them.
There is an urgent need for automated vulnerability remediation in continuous fuzzing pipelines to both ease the developers' workload and minimize the window of vulnerability exposure.


Recent advancements in generative AI and LLM agents have shown promise in autonomous vulnerability remediation in programs ~\cite{zhang2024acr, ruan2024specrover, yang2024sweagent, xia2024agentless}. 
These LLM agents are designed for general software engineering tasks, including bug fixing and feature development. 
They operate in real-world scenarios where tasks are described by users in natural language. 
Using the task description and the software codebase as inputs, the agents generate code modification suggestions to fulfill the specified requirements.
Since repairing security vulnerabilities is a specialized software engineering task, we hypothesize that with appropriate adaptation, general-purpose LLM agents for software engineering can be repurposed for this task.
These repurposed agents can potentially be integrated into existing vulnerability detection pipelines such as fuzzing, where they can provide the remediation after detection and complete the software protection cycle.

\paragraph{\tool} In this paper, we present a large scale real-world study on using LLM agents for security vulnerability repair. To enhance the realism of our effort, we use as dataset the OSS-Fuzz projects, which seek to enhance the state of practice of open source security \cite{ossfuzz}.
We repurposed the open-source LLM agent \acr~\cite{zhang2024acr, ruan2024specrover} to repair security vulnerabilities, and implemented a version named \tool (i.e. \acr for security).
With the vulnerability report and an exploit input produced by a fuzzing campaign, \tool autonomously generates patches that fix the detected vulnerability.
In the process of adapting LLM agents for vulnerability repair, we identified that one of the main challenge was the insufficient information contained in the auto-generated vulnerability report.
Unlike human-written issue report for general software engineering tasks, vulnerability reports are often auto-generated by the fuzzer and only contain information like the bug type and crash stacktrace.
To enrich the vulnerability report, we extract dynamic call graph information from the exploit input found by fuzzing, which is then used to augment the report generated by the fuzzer.
In addition, we perform a type-based analysis at the program locations identified as faulty by the agent, and use the additional type information to improve the compilation rate of generated patches.
With these adaptation, we built \tool which can autonomously repair security vulnerabilities detected in a fuzzing pipeline.

To evaluate the efficacy of \tool in a real-world setup, we conducted experiments on real C/C++ vulnerabilities previously detected by \ossfuzz. 
Each detected vulnerability comes with an exploit input that resulted in a crash from sanitizers (e.g. AddressSanitizer~\cite{asan}, MemorySanitizer~\cite{msan}), and the crash report generated by the sanitizer.
Experiments on 588 real-world vulnerabilities from a previously curated dataset~\cite{mei2024arvo} show that \tool can repair 52.4\% of these vulnerabilities by resolving the crash from the exploit input.
We also comparatively study the efficacy of other deep learning or LLM-based systems in this realistic vulnerability repair scenario.
Firstly, we apply a general-purpose LLM coding agent directly to the vulnerability repair setting and observe a lower repair efficacy.
Secondly, we evaluate the state-of-the-art deep learning based vulnerability repair system \vulmaster~\cite{zhou24vulmaster}
in the \ossfuzz dataset.
Existing deep learning based vulnerability repair techniques often make strong assumptions such as the perfect fix location is provided at either function- or line-level. 
This localization assumption is too strong for real-world vulnerability repair, and we observe a low repair efficacy when they are used in a realistic repair setup.
Furthermore, current evaluation of vulnerability repair tools often focus on how closely the generated patches match the developer's patch, using metrics like exact match or similarity scores such as BLEU~\cite{papineni2002bleu} or CodeBLEU~\cite{ren2020codebleu}.
However, we find that these metrics may not accurately reflect the true efficacy of the repairs.
Therefore, it is important to assess vulnerability repair systems on datasets with executable inputs in future research.

In summary, our contributions are as follows:
\begin{itemize}
    \item We explore the feasibility of adapting general-purpose LLM programming agents for the repair of security vulnerabilities.
    We integrate call graph information and type-based analysis to provide richer context for LLM agent-based vulnerability repair, resulting in improved patch quality.
    Our approach is implemented as a new agent \tool which is specialized for security vulnerability repair.
    \item We conduct an empirical study on the use of LLM agents to repair real-world security vulnerabilities identified by the industrial fuzzing service \ossfuzz.
    Our findings indicate that leveraging LLM agents for vulnerability remediation is a promising approach to complement existing detection pipelines and complete the software protection life cycle.
    \item We present empirical evidence suggesting that similarity scores may not accurately measure the effectiveness of vulnerability repair systems. To better evaluate learning-based vulnerability repair systems, future research should consider test-based validation methods.
\end{itemize}

%% file: ossfuzz.tex
\section{The \ossfuzz Case Study}
We discuss background on the \ossfuzz project and LLM agents for software engineering. 

\label{sec:background}
\subsection{Overview of OSS-Fuzz project}
\label{sec:oss_fuzz}
\begin{figure}[t]
\includegraphics[width=0.55\linewidth]{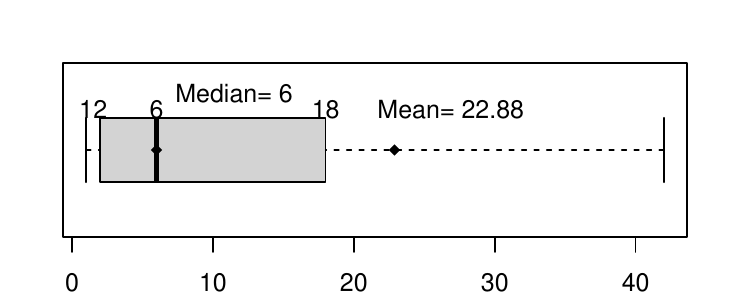}
\caption{The boxplot represents the distribution of the number of bugs identified through fuzzing since 2016. The median and mean values are highlighted in the diagram with some outlier data points are removed for better presentation.}
\label{fig:no_vuls}
\end{figure}

Fuzz testing \cite{fuzzing} is a popular method for detecting software security vulnerabilities, via a biased random search over the domain of program inputs.
Launched by Google in 2016, \ossfuzz is an open-source initiative designed to continuously detect security vulnerabilities across more than 1250 open-source software projects. 
The participating projects provide a fuzzing harness to test specific API functions. 
OSS-Fuzz monitors the reliability of software projects' repositories
by continuously testing them with a wide range of fuzzers (e.g. AFL++, libfuzzer, Hongfuzz) and sanitizers (e.g. AddressSanitizer~\cite{asan} and UndefinedBehaviorSanitizer~\cite{ubsan}). 
It automatically reports any crashes identified by the fuzzers and periodically verifies whether the project has resolved the reported issues. 
As of September 2024, the \ossfuzz cluster has discovered over 12,000 bugs across all projects. 
As shown in \autoref{fig:no_vuls}, on average, \ossfuzz has reported 22 bugs for each participating project, with some projects such as \emph{ffmpeg} having over 400 bugs reported.
In this work, we focus on vulnerability remediation for C/C++ vulnerabilities detected by \ossfuzz.

\subsection{LLM Agents for Software Engineering}
\label{sec:sut}

Recent advances in Large Language Models' (LLMs) context windows have significantly improved their ability to process complex text sequences. 
This enhancement, combined with their capacity for task planning, has led to the development of agent-based systems designed to tackle a broad spectrum of problems.
The model solves the task by decomposing it into smaller, manageable steps that another model or tool can handle.

One area where notable success was demonstrated is software engineering. 
Here, an agentic system is provided with a natural language description of a task, such as issue descriptions in software repositories like GitHub. 
The issue can describe a bug or new features to be added to a codebase.
To solve the issue, the LLM can invoke external tools, allowing it to interact with the environment and gather more data before presenting a solution in the form of a patch. 
These tools encompass actions fundamental to software engineering, such as Abstract Syntax Tree (AST) search, e.g., getting a function or class definition, file system navigation, and executing commands such as compiling the project or running the test suite. 
By integrating such tools, the agent can analyze the codebase, invoke tools to gather additional information about the failure, and make modifications while keeping track of the original task.
Examples of state-of-the-art LLM Agents are \acr~\cite{zhang2024acr}, \sweagent~\cite{yang2024sweagent}, and \agentless~\cite{xia2024agentless}.

The approach and the system prototype used in our research is the autonomous program improvement agent \acr~\cite{zhang2024acr},\cite{ruan2024specrover}.
The work of \acr shows that natural language specifications such as GitHub issue reports are sufficient for driving an autonomous agent to fix bugs or add features. 
\acr coordinates a group of agents to achieve context retrieval, following which a patch writer (known as \emph{Patch Agent}) will produce the final patch.
If a test suite is available, \acr will execute the test suite to validate whether the proposed patch satisfies the test suite. 
When tests are not available,
\acr employs a \emph{Test Agent} to construct a test case reproducing the issue.
By delegating specific tasks to dedicated agents within the workflow, the system can efficiently utilize LLMs' strengths in natural language understanding and code generation.

\textbf{Significance.}
Repairing vulnerabilities detected by fuzzing is vital for enhancing software security and reliability, as evidenced by the efforts from both software engineering research~\cite{zhang2022vulfix, chen2021neural, gao2021beyond} and industry~\cite{keller2024patching}.
According to a recent study by Mei et al.~\cite{mei2024arvo}, the number of vulnerabilities identified by \ossfuzz is growing steadily despite the gap between reproducible vulnerabilities and their fixes, posing a significant security risk.
Furthermore, the rising number of unpatched vulnerabilities over time implies that some vulnerabilities might not receive immediate attention.
Therefore, it is essential to propose reliable solutions for vulnerability remediation.

%% file: methodology.tex
\section{\tool}
\label{sec:method}

To study whether LLM agents for general software engineering tasks can be specialized for vulnerability remediation, we adapted the 
open-source agent \acr for security vulnerability repair.
In this section, we first provide an overview of \acr, and subsequently discuss how we repurposed it for vulnerability repair.

\subsection{Overview of \acr}
\label{sec:autocoderover}

\begin{figure}[t]
     \centering
     \includegraphics[width=\linewidth]{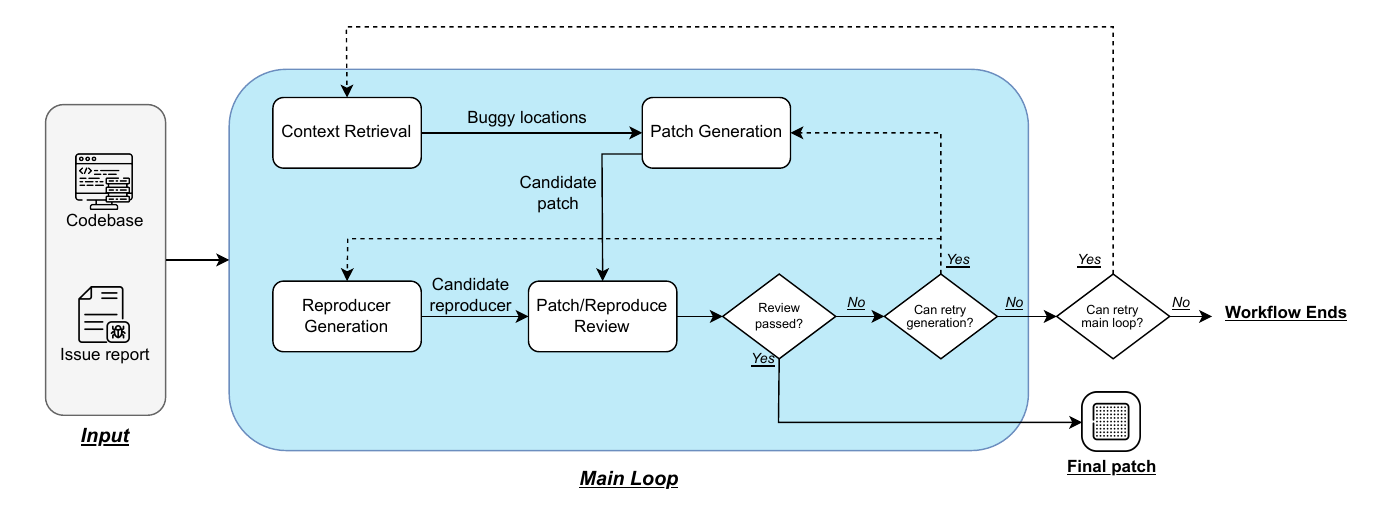}
     \caption{Workflow of \acr for resolving GitHub issues. Dotted lines are back edges in the workflow.}
     \label{fig:workflow-acr}
\end{figure}

\acr~\cite{zhang2024acr} is an LLM agent designed for software engineering tasks like bug fixing and feature addition.
It aims to resolve software engineering issues in a realistic setup, where only a natural-language description of the issue/requirement is available.
One such setup is GitHub issues, in which users submit bug reports or feature addition requests to a software project.

\begin{figure}[t]
     \centering
     \begin{subfigure}{0.45\textwidth} 
        \centering
         \includegraphics[width=\textwidth]{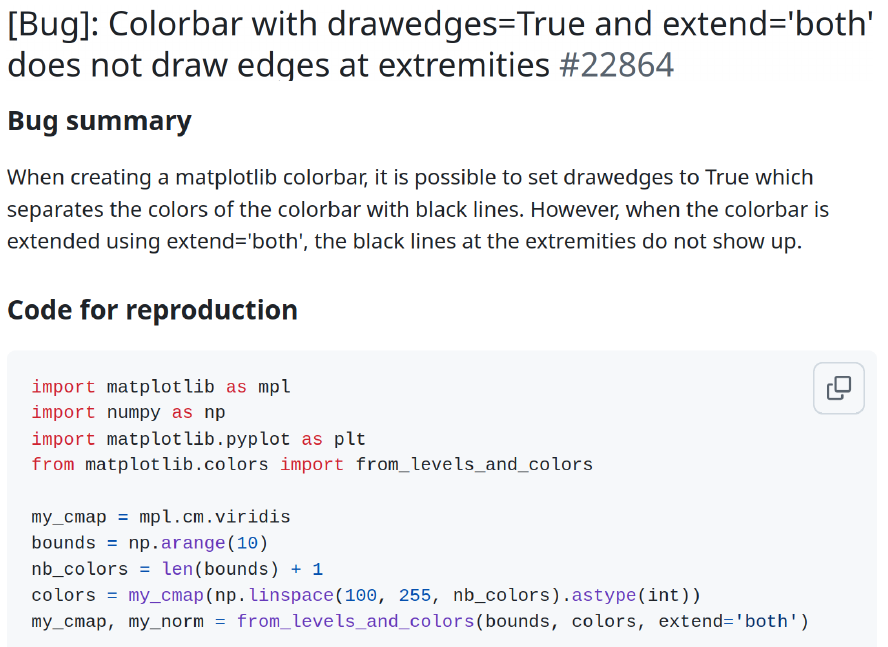}
         \caption{An example GitHub issue.\protect\footnotemark}
         \label{fig:github_issue}
     \end{subfigure}
     \begin{subfigure}{0.48\textwidth} 
            \centering
             \input{figs/sanitizer_report}
            \caption{Example of sanitizer report for Kamailio-38065\protect\footnotemark.}
            \label{fig:sanitizer_report}
     \end{subfigure}
     \caption{Examples of GitHub issues and sanitizer-generated reports.}
\end{figure}

\autoref{fig:workflow-acr} illustrates the workflow of \acr in resolving GitHub issues. 
Given a codebase \textit{C} and a natural-language (NL) issue report \textit{R}, \acr autonomously produces a patch that aims to resolve the issue described in \textit{R}.
From the issue report \textit{R}, \acr begins the main loop with \textit{context retrieval} and \textit{reproducer generation}.
Since an issue typically only contains NL descriptions and no executable test to reproduce the issue, \acr first attempts to generate a candidate reproducer test for the given issue.
This reproducer test serves as an additional specification for the patch generation later on.
Other than the reproducer test, \acr also starts the \textit{context retrieval} stage from the issue report \textit{R}.
The goal of context retrieval is to extract code snippets relevant to the issue \textit{R} from a large codebase, enabling the LLM to better understand the issue in relation to the code.
\acr performs context retrieval by designing a set of program structure-aware \textit{search APIs}, and allowing the LLM to interact with a local codebase through these APIs.
Specifically, \acr provides a set of API specifications such as \texttt{search\_class(...)}, \texttt{search\_method\_in\_class(...)}, etc., to the LLM.
The LLM is instructed to invoke some of these APIs with appropriate arguments based on what is described in the issue report \textit{R}.
For example, given the example issue shown in \autoref{fig:github_issue},  
the LLM would likely invoke the API \texttt{search\_class("Colorbar")} to obtain more context about this class.
Upon receiving such an API request, the backend of \acr searches for the actual code/signature of the class \texttt{Colorbar} from an Abstract Syntax Tree (AST) representation of the codebase, and returns the code/signature back to the LLM.
This process of invoking search APIs and enriching code context happens iteratively, until the LLM deems that the current code context is sufficient for understanding the issue.
At the end of the context retrieval stage, the LLM decides on a few \textit{buggy locations} from the current code context. 
These buggy locations are provided to a \textit{patch generation} module to craft candidate patches that aim to resolve the issue.

After a candidate patch is generated, \acr attempts to examine whether it resolves the issue in a \textit{review} module.
If the patch is deemed to resolve the issue, the workflow ends with it being the final patch.
Otherwise, a natural-language ``suggestion'' on how to improve the current patch is sent back to the patch generation module to iteratively improve the patch.
A natural way to decide whether a patch resolves the issue is to execute the reproducer test on the patched program. 
The \textit{review} module in \acr takes in both the generated candidate patch and test, execute them, and decide on (1) whether the test successfully reproduce the issue, and (2) whether the patch successfully resolves the issue.
Both the test and patch are subject to iterative refinement since both of them are generated by the LLM and can be incorrect.
If no acceptable patches were generated after several rounds of review, \acr goes back to the context retrieval stage to re-discover buggy locations and a new set of patches.
This process continues up to a pre-defined number of rounds.

\footnotetext[1]{Issue \#22864 from the matplotlib project. \url{https://github.com/matplotlib/matplotlib/issues/22864}}
\footnotetext[2]{Bug \#38065 from the Kamalio project, detected by \ossfuzz. \url{https://bugs.chromium.org/p/oss-fuzz/issues/detail?id=38065}}

\subsection{\tool for Security Vulnerability Repair}
\label{sec:coderover-s}

The \acr workflow presented in Section~\ref{sec:autocoderover} is designed for resolving software engineering issues with natural language descriptions.
We next discuss the adaptation of \acr into the context of repairing vulnerabilities detected by fuzzing campaigns such as in \ossfuzz.
This adaptation results in an LLM agent for security vulnerability repair, which we call \tool.

We observe two main differences between the scenario of resolving GitHub issues and repairing vulnerabilities detected by fuzzers:

\begin{enumerate}
    \item Although GitHub issues sometimes contain steps to reproduce, these steps are usually provided by the user and may not be fully executable in the environment where the repair workflow is run.
    Therefore, agents targeting GitHub issues often employ an LLM to extract an executable reproducer test from the issue report.
    In contrast, vulnerabilities detected by fuzzing always come with a Proof-of-Vulnerability exploit input. 
    This exploit input is generated by the fuzzer to prove the existence of the vulnerability, and can be used to reproduce the vulnerability in the fuzzing environment (assuming the vulnerability has been confirmed and is not flaky).
    
    \item GitHub issue reports are usually human-written and contain natural-language descriptions of the issue.
    These natural language descriptions typically cover the program components that are related to the issues and some additional details, as shown in the example in \autoref{fig:github_issue}.
    In contrast, vulnerabilities found by fuzzers usually come with a bug report generated by sanitizers such as AddressSanitizer.
    An example of sanitizer-generated report is shown in \autoref{fig:sanitizer_report}.
    These auto-generated vulnerability reports contain the error type and a stack trace of executing the exploit input, but does not contain more elaborated natural language description of the root cause.
\end{enumerate}

\begin{figure}[t]
     \centering
     \includegraphics[width=0.9\linewidth]{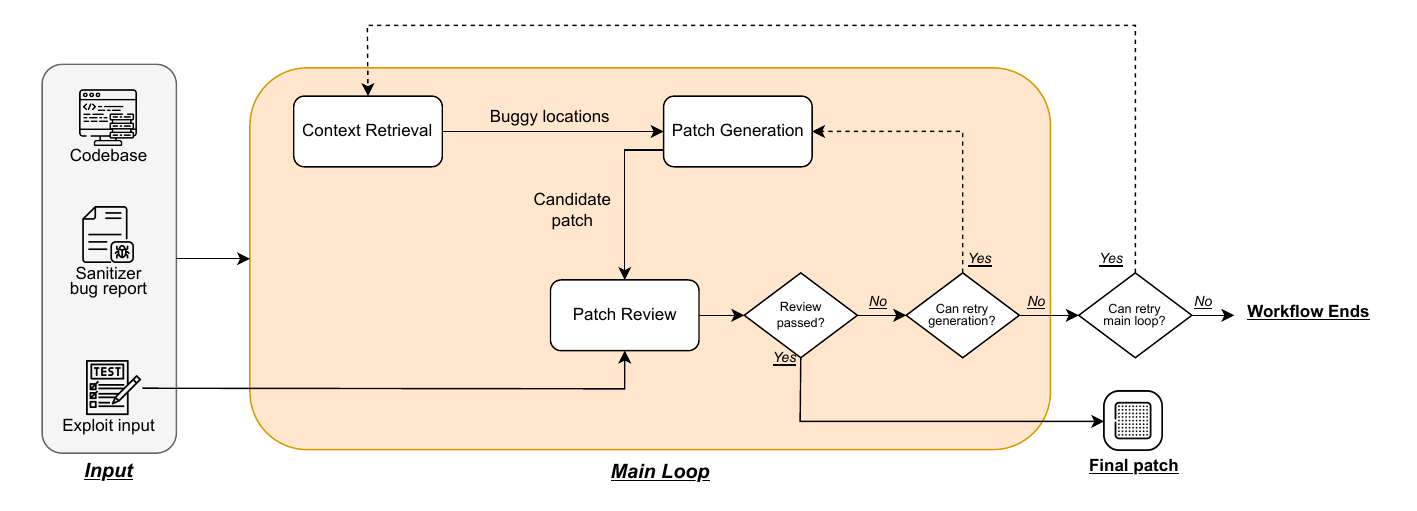}
    \caption{Workflow of \tool for repairing security vulnerabilities.}
     \label{fig:workflow_coderover-s}
\end{figure}

Based on these observations, we propose \tool, an LLM agent built on top of \acr but tailored for security vulnerability repair.
\autoref{fig:workflow_coderover-s} presents the workflow of \tool for repairing vulnerabilities found by fuzzing.
Compared to \acr, \tool additionally takes in the fuzzer-generated exploit input, and uses this input for candidate patch validation within the workflow.
Since the exploit input can reliably reproduce the vulnerability (c.f. Observation (1)), \tool does not attempt to generate reproducer inputs from the bug report.
This exploit input is used during the \textit{Patch Review} stage, in which the exploit input is executed against the generated candidate patches to examine whether the patches can resolve the crash from the exploit input.
Similar to \acr, if a candidate patch does not pass the review, \tool iteratively refines the generated patch until a pre-defined limit of retries is reached.
In contrast to \acr where the review judges the correctness of both test and patch, the review stage in \tool assumes the test is correct and examines the patch correctness based on the test execution result.
The rest of the workflow follows that of \acr: if the candidate patch is rejected at the review stage, retries will happen at both the patch generation and context retrieval level, depending on the pre-defined retry limits.
\tool will output a final patch if the patch passes the review stage, or output the best patch so far (based on heuristics such as whether the patch can be compiled) if no patches have passed the review.

We now elaborate in the next subsection, how to tackle the challenge of creating security patches from sanitizer reports.

\subsection{Vulnerability Repair from Sanitizer Report}\label{sec:bug_report}

In addition to adapting the agent workflow to security vulnerability repair, another challenge is to augment the bug report since the auto-generated report contains limited information (c.f. Observation (2)). 
Auto-generated bug reports from fuzzing are typically the output from sanitizers when executing the exploit input on the buggy program.
Security sanitizers instrument the original program, and turns buggy executions that exhibit certain types of vulnerabilities into program crashes.
For example, AddressSanitizer~\cite{asan} and UndefinedBehaviorSanitizer~\cite{ubsan} are widely used sanitizers that turn memory errors and undefined behaviors into crashes.

\autoref{fig:sanitizer_report} shows an example of bug report generated by AddressSanitizer for a vulnerability found by \ossfuzz in the Kamailo project.
The sanitizer report includes the bug type (e.g. heap-buffer-overflow) and the stack trace when the crash happens.
Though the bug type captures the requirement for the agent (i.e. ``fix the heap-buffer-overflow in the program''), the crashing stack trace may not provide sufficient context for the agent to perform context retrieval.
Agents like \acr and \tool explore the codebase and extract the code context based on the information in the issue/bug report.
They may invoke search APIs such as \texttt{search\_method\_in\_file(...)} to initiate the context retrieval process and progressively search for other code elements as needed.
However, the sanitizer report only includes the lines and functions in the crash stack trace, which is only a small part of the entire execution and may not serve as a good starting point for context retrieval.
For example, the bug Kamailio-38065 shown in \autoref{fig:sanitizer_report} was patched by the developer with the changes shown in Figure~\ref{fig:kamailo_developer_patch}.
The developer patch modifies the \verb|skip_name| function, which does not appear in the stack trace but is invoked in other parts of the execution.
It could be challenging for the agent to use the sanitizer report as the starting point of context retrieval and navigate to this function in the codebase.
This may restrict the retrieved context and affect the quality of the final patch.
To address the challenge of limited information in the sanitizer report, we enrich the auto-generated report with additional context of the buggy execution.
Specifically, we take advantage of the available exploit input, and generate a dynamic call graph from the execution of the exploit input.
This dynamic call graph is used to augment the sanitizer report and provides more contextual information for the agent to navigate the codebase.

\begin{figure}[t]
    \centering 
\begin{minipage}{0.7\linewidth} 
  \begin{lstlisting}[style=diffstyle, basicstyle=\tiny]
index 8c6ebdd6bb..345167022f 100644
--- a/src/core/parser/contact/contact.c
+++ b/src/core/parser/contact/contact.c
@@ -147,10 +147,10 @@ static inline int skip_name(str* _s)
     return 0;
     }
- if (*p == ':') {
+ if (*p == ':' || *p == ';') {
     if (last_wsp) {
-        _s->s = last_wsp;
         _s->len -= last_wsp - _s->s + 1;
+        _s->s = last_wsp;
     }
     return 0;
  }
\end{lstlisting}
\end{minipage}
\caption{The developer's patch for fixing Kamailio-38065.\protect\footnotemark}
\label{fig:kamailo_developer_patch}
\end{figure}

\footnotetext[3]{\url{https://github.com/kamailio/kamailio/commit/20db418f1e35f31d7a90d7cabbd22ae989b7266c}}

\paragraph{Dynamic Call Graph Construction}
To construct a dynamic call graph from the buggy execution, we instrument the buggy program during compile time to insert hooks at every function entry and exit points. 
These hooks record the memory addresses of the functions, as well as the calling relationships between callers and callees.
The instrumented program is then executed with the exploit input to trigger the vulnerability. 
During the execution, the function entry/exit hooks log an edge list comprising pairs of function call addresses. 
Following this procedure, we map the memory addresses to their original function names and source code locations (e.g. filename and line number).
In practice, we utilize \emph{addr2line}~\cite{addr2line}, \emph{gdb}~\cite{gdb}, and \emph{nm}~\cite{nm_tool} to accomplish such mappings.
Since the call graph size of real software project executions can be substantial, we performed the following post-processing to enhance the clarity of the information extracted from call graph:

\begin{itemize}[leftmargin=*]
    \item Filtering functions: Invocation of standard libraries and internal functions that are not part of the codebase were removed, to focus the analysis on functions directly involved in the vulnerability.
    For example, calls to standard C library functions like \texttt{malloc()}, \texttt{free()}, and \texttt{memcpy()}, or logging functions such as \texttt{log\_debug()}, were filtered out.
    \item Demangling function Names: For languages that utilize name mangling (e.g. C++), the function names were de-mangled to the original, human readable identifiers in the source code. 
    For example, the name \texttt{\_ZL28demangling\_terminate\_handlerv} would be de-mangled to \\ \texttt{demangling\_terminate\_handler}.
\end{itemize}

\begin{figure}[t]
     \centering
     \includegraphics[width=0.6\linewidth]{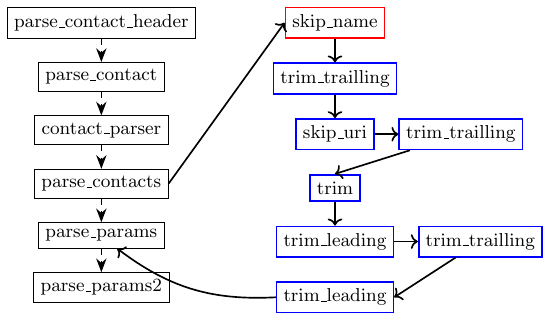}
     \caption{An example of enhancing the bug report using a dynamical call graph. The dashed lines represent the order of function calls on stack trace and the solid lines augment them to show  the actual dynamical call graph. The red colored function call is the fix location selected by the developer.}
     \label{fig:cg_diagram}
\end{figure}

\autoref{fig:cg_diagram} shows an example of the constructed dynamic call graph for the bug Kamailo-38065.
The red and blue-colored function calls display the part of the dynamic call graph beyond the crash stack trace. 
These function calls serve as additional starting points for the agent to explore the codebase.
To make the call graph available to the LLM, we concatenate the list of additional function calls to the sanitizer bug report as ``other functions executed by the bug-triggering input''.
The functions invoked closer to the crash location are presented earlier in this list.
The additional list of function calls enriches the auto-generated bug report from sanitizers, and 
provides more context for code retrieval in \tool.

\subsection{Type-assisted Patch Generation}
\label{sec:repair_prompt}

The context retrieval stage of \tool outputs a list of buggy locations (e.g. functions).
With the code snippets at the buggy program locations, the patch generation stage attempts to craft patches that fixes the vulnerability.
However, a generated patch may not be always compilable.
This is because the code of the buggy function itself may not contain the necessary patch ingredient.
For example, if the type definition of a struct variable \texttt{s} is not within the buggy function, the LLM may hallucinate some field names of \texttt{s} and use those names in the patch, which will make the patch not compilable.
A straightforward solution is to provide the entire file content around the buggy functions to the LLM.
However, code files can be large in real-world C/C++ projects, and may not fit in the context window of the LLM.
Even if the entire file fits within the context window, the relevant type definitions might be absent, as they could be defined in separate header files.

\begin{figure}[t]
     \centering
     \input{figs/repair_prompt}
     \caption{Augmented repair prompt using context information.}
     \label{fig:repair_prompt}
\end{figure}

To ensure the relevant context is present for patch generation, we introduce 
a type-assisted patch generation prompt that includes all existing variables and their types in the scope of the buggy function.
To compose such a prompt, we parse the C/C++ source files to capture important language constructs such as structs, classes, typedefs, and enums. 
Using this information, we construct a type graph representing the relationships between types and variables. 
This step handles complex type relationships, including type aliases through typedefs, while maintaining information about structs, unions, and enums.
This constructed type graph is added to the patch generation prompt so that the LLM can refer to these type information when writing a patch, thereby improving the compilation rate of the patches.
An example patch generation prompt for the example vulnerability Kamailio-38065 is shown in \autoref{fig:repair_prompt}.

\subsection{Language-specific Features}

Since the original \acr was implemented for resolving GitHub issues in Python repositories, we implemented additional support for C/C++ projects in \tool.

\paragraph{Parsing of C/C++ code}
\acr and \tool parse the project source code into an AST to facilitate future code retrieval from the search APIs.
We implemented the parsing of C/C++ source files for the search index of \tool using \emph{tree-sitter}~\footnote{\url{https://tree-sitter.github.io/tree-sitter/}}.
To improve the search accuracy for identifying entities such as functions and structures, we process the standard C/C++ code patterns and some uncommon ones, such as function definitions through macros and structures defined via typedefs.

\paragraph{Search APIs}
We also modify the search API in \tool to explore the codebase with a higher success rate.
One notable feature of C/C++ is forward declarations, i.e., defining a function or class without providing an implementation for it.
This leads to implementations scattered across the codebase.
To handle this, we constrain the search API by removing tools that  may confuse the model when the search is unsuccessful, such as \verb|seach_class_in_file|. 
Furthermore, the stack trace in the sanitizer report does not contain information about the structure of qualified identifiers in C++.
For example, there can be a method identifier of the form \verb|A::B|, but it is unclear whether \verb|A| refers to a class or a namespace.  
To handle this, we modify the search functionality to search across namespaces, classes, and structures when searching for a method containing such an identifier.

%% file: figs/sanitizer_report.tex
\begin{tcolorbox}[size=small,boxrule=0pt,title={\small AddressSanitizer Report},width=\linewidth,]
\tiny{\color{red}==24==ERROR: AddressSanitizer: heap-buffer-overflow on address 0x60e000000293 ... }\\
{\color{red}READ of size 1 at 0x60e000000293 thread T0}\\
\textbf{\#0} 0xe7763d in \textit{q\_memchr} src/core/parser/../ut.h:422:7\\
\textbf{\#1} 0xe771e8 in \textit{parse\_quoted\_param} src/core/parser/parse\_param.c:305:14\\
\textbf{\#2} 0xe7175a in \textit{parse\_param\_body} src/core/parser/parse\_param.c:450:6\\
\textbf{\#3} 0xe6b2d8 in \textit{parse\_param2} src/core/parser/parse\_param.c:496:13\\
\textbf{\#4} 0xe6d274 in \textit{parse\_params2} src/core/parser/parse\_param.c:599:10\\
\textbf{\#5} 0xe6ce56 in \textit{parse\_params} src/core/parser/parse\_param.c:561:9\\
\textbf{\#6} 0xeb16b2 in \textit{parse\_contacts} src/core/parser/contact/contact.c:243:8\\
\textbf{\#7} 0xe4a638 in \textit{contact\_parser} src/core/parser/contact/parse\_contact.c:55:7\\
\textbf{\#8} 0xe49405 in \textit{parse\_contact} src/core/parser/contact/parse\_contact.c:84:6\\
\textbf{\#9} 0x87e4f4 in \textit{parse\_contact\_header} src/core/select\_core.c:234:9\\
...
\end{tcolorbox}

%% file: figs/repair_prompt.tex
\begin{tcolorbox}[size=small,boxrule=0pt,title=Repair Prompt,width=0.8\textwidth,fonttitle=\footnotesize,]
\footnotesize
To ensure the patch is compilable, please use only existing variables at the specified bug locations.\\
Here's a list of available variables and their types:\\
...


\begin{verbatim}
variables in method: parse_param_body
Variables declarations:
- name: p , type:  param_t*
       typedef: param_t originaltype:struct param ...
- name: _s , type:  str*
       typedef: str original_type:* json_key
       ...
- name: separator , type:  char
- name: _c , type:  pclass_t
       typedef: pclass_t original_type:enum pclass ...
\end{verbatim}
When writing your patch, make sure to:\\
    1. Use variables in a way that's consistent with their types.\\
    2. Do not introduce imaginary variables that do not exist within the existing code snippet or the provided context.\\
Write a patch for the vulnerability, based on the relevant code context.
First explain the reasoning, and then write the actual patch.
When writing the patch, remember the following:\\
 - You do not have to modify every location provided - just make the necessary changes.\\
 - Other than the vulnerability to fix, your patch should preserve the program functionality as much as possible.
...
\end{tcolorbox}

%% file: evaluation.tex
\section{Evaluation}
\label{sec:eval}

To study the effectiveness of automated tools in real-world vulnerability remediation, we evaluate various systems on real vulnerabilities detected by \ossfuzz.
We aim to examine the efficacy of LLM agents and learning-based vulnerability repair systems in a realistic vulnerability repair scenario.
Specifically, we would like to answer the following research questions:
\begin{itemize}[leftmargin=*]
    \item RQ1: What is the efficacy of \tool in repairing real-world security vulnerabilities compared to other tools?
    \item RQ2: What are the strength and weakness of \tool in vulnerability repair?
    \item RQ3: What are the suitable metrics for patch evaluation in security vulnerability repair?
\end{itemize}

\paragraph{Benchmark Selection.}
To study the effectiveness in repairing vulnerabilities detected by \ossfuzz, we utilize the recently introduced reproducible benchmark dataset, ARVO~\cite{mei2024arvo}.
The ARVO dataset contains 5,001 C/C++ vulnerabilities detected by \ossfuzz across 273 projects, and each vulnerability comes with an environment to rebuild the buggy project and a bug-triggering exploit input. 
Since a large number of vulnerabilities have been discovered by \ossfuzz in the past years and are contained in the ARVO dataset, we employ a widely adopted sampling strategy~\cite{baltes2022sampling} to construct a subset of vulnerabilities as our dataset.

Specifically, we use the 5,001 vulnerabilities in the ARVO dataset as the population, and apply a sampling method to create a representative subset. 
After applying a sampling methodology with a 99\% confidence level and a margin of error of 5\%, we select an initial subset of 588 bug samples for further in-depth analysis.
From the initial subset, we repeatedly filter out bugs based on the following criteria and re-sampled more bugs to eventually obtain a final subset of 588 vulnerabilities: (1) the reported vulnerability should be reproducible in our experiment environment; (2) the compilation time of the buggy project should be under 15 minutes. 
This is to ensure the builds during the repair/validation process can finish within a reasonable time.

\begin{figure}[ht]
    \centering
    \begin{subfigure}{0.42\textwidth} 
     \includegraphics[width=\textwidth]{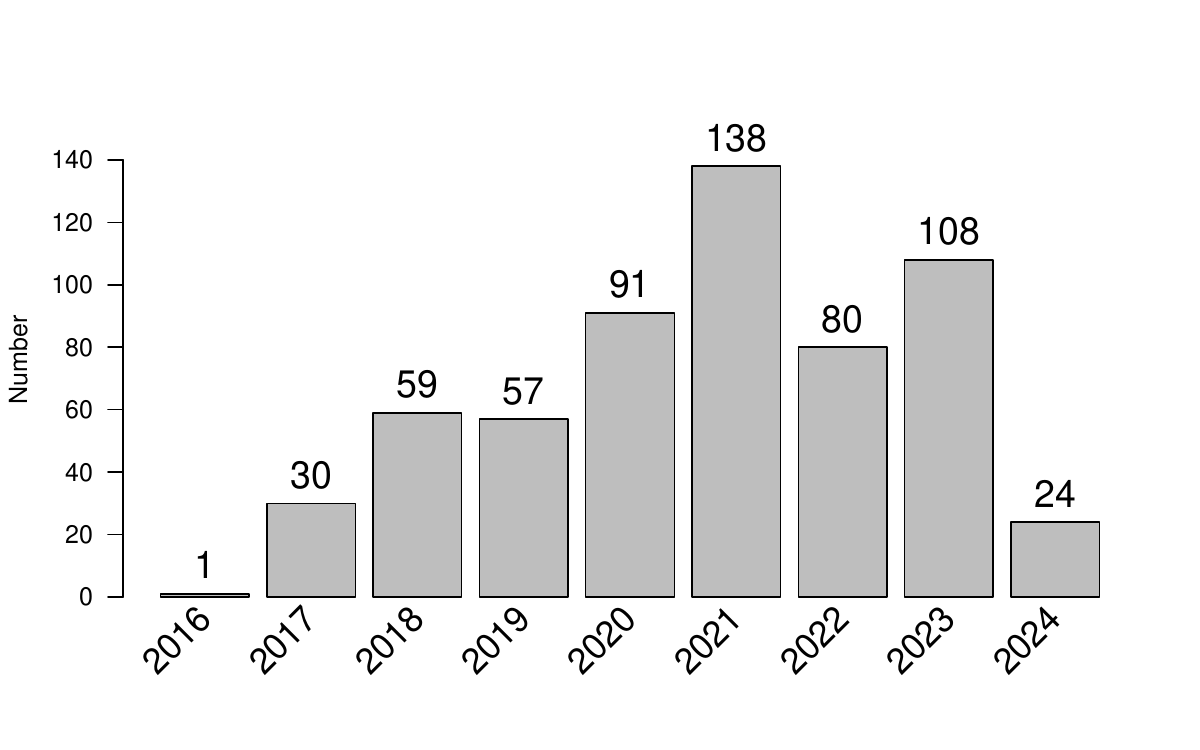}
    \end{subfigure} 
    \begin{subfigure}{0.55\textwidth}  
    \includegraphics[width=\textwidth]{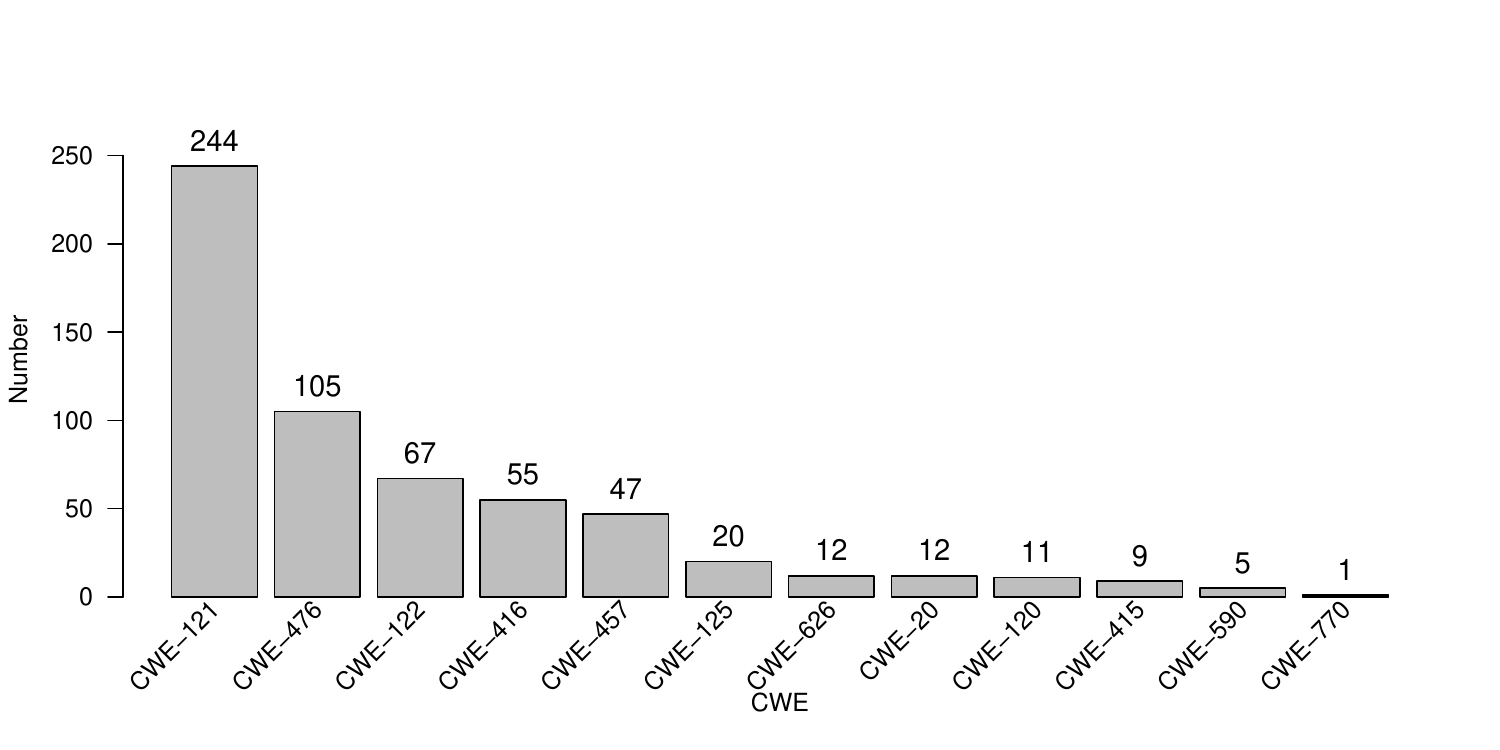}
    \end{subfigure}
    \caption{The number of bugs detected by years and CWEs, respectively.}
    \label{fig:bugs_distribution}
\end{figure}

\autoref{fig:bugs_distribution} presents the distribution of vulnerabilities in our dataset based on the year and CWE type they were detected.
The most common CWE type are buffer overflows (CWE-121 heap-based overflow and CWE-122 stack-based overflow), segmentation faults (CWE-476), use-after-free (CWE-416), and use of uninitialized value (CWE-457).

\paragraph{Baseline tools}
We compare the efficacy of \tool with two baseline tools:
\begin{enumerate}[leftmargin=*]
    \item \agentless~\cite{xia2024agentless}: \agentless is the state-of-the-art LLM-based agentic system designed for general-purpose software engineering tasks, as of August 2024~\cite{swebench-verified}.
    \agentless employs a fixed two-phase workflow of localization and repair. The localization phase happens on multiple granularities, such as classes, methods, and file names.
    The repair phase generates multiple patches at the identified fix locations, and a patch selection process is employed to pick the final patch.
    \agentless was originally designed for Python codebases.
    To use \agentless for the C/C++ projects in \ossfuzz, we have re-implemented the project structure generation, class/function parsing, and edit location mapping for C/C++.
    Moreover, some prompts are adjusted from Python filenames or Python code snippet to that of C/C++.
    We use the sanitizer-generated bug report for each vulnerability as its input.
    For patch selection, since there is an exploit input available for each vulnerability, we use the following rule to select the final patch:
    \emph{A plausible patch is strictly preferred to implausible one, and an implausible patch is strictly preferred to the one with compilation issue}.
    If there are multiple plausible patches, the first one is picked as the final patch.

    \item \vulmaster~\cite{zhou24vulmaster}: \vulmaster is a learning-based vulnerability repair tool. 
    \vulmaster finetunes a CodeT5 model using vulnerability repair data such as CWE identifiers, vulnerability descriptions, exemplar repairs and the vulnerable program fragment to be repaired. 
    The output of \vulmaster is the repaired version of the vulnerable fragment.
    Learning-based vulnerability repair tools like \vulmaster usually assume \textit{perfect localization} where the vulnerable program statement/expression is given.
    They require the input program fragment to mark the to-be-fixed statement/expression with two special tokens \verb|<start-vul>| and \verb|<end-vul>|.
    To use \vulmaster in a realistic vulnerability repair scenario, we employ the standard Spectrum-based Fault Localization (SBFL) to obtain a few candidate fix locations, and run \vulmaster over these locations.
    This SBFL process takes in the exploit input as the failing test, and other non-crashing inputs generated from the \ossfuzz fuzzing campaign as the passing tests, and computes a list of suspicious locations using the Ochiai metric~\cite{ochiai}.
    We use the top-five suspicious locations (line-level) from SBFL as the fix locations, and use \vulmaster to generate a patch at each of these locations.
    Since the trained model of \vulmaster is not publicly released, we follow the same procedure to finetune the CodeT5 model using the same training dataset used in the original paper.
    Since there are multiple patches produced (one at each location), we select the final patch by following the same process as that in \agentless (i.e. always select the plausible patches first).
\end{enumerate}

For each vulnerability in the setup, the inputs to \tool and \agentless are:
\begin{enumerate}[leftmargin=*]
    \item An exploit input that triggers the vulnerability.
    \item The sanitizer-generated bug report for this vulnerability. This is obtained by executing the exploit input on the program built with sanitizers.
\end{enumerate}

For \vulmaster, we follow its original setup~\cite{zhou24vulmaster} and give it the following inputs:
\begin{enumerate}[leftmargin=*]
    \item CWE identifier extracted from the sanitizer-generated report.
    \item A few exemplar fixes for the corresponding CWE type, extracted from the CVEFixes dataset~\cite{bhandari2021cvefixes}.
    \item The function containing the fix location identified from SBFL.
\end{enumerate}

\paragraph{Tool Setup}
For agent systems (\tool and \agentless) used in our evaluation, we used the OpenAI \texttt{gpt-4o-2024-08-06} as the backend LLM.
We set the model temperature to 0.2, following previous works~\cite{zhang2024acr}.
For \tool, we set the maximum rounds of main loop to be 3, following the same experimental setup in \cite{ruan2024specrover}.

We evaluate the final patch generated by a tool by  
applying it to the buggy program, compiling the patched program, and executing the exploit input on the patched program. 
We then classify the patches into the following categories:
(1) No Patch (\textbf{NP}): A patch is not generated by the tool.
(2) Compilation error (\textbf{CE}): A patch is generated, but caused compilation errors when building the program.
(3) Implausible (\textbf{IP}): The generated patch compiles without errors, but the original exploit input still triggers the vulnerability.
(4) Plausible (\textbf{P}): The patch compiles successfully, and the original exploit input can no longer trigger the bug.

\subsection{RQ1: Repair Efficacy of Various Tools}

In this research question, we aim to evaluate the capabilities of the LLM- and learning-based repair systems in repairing vulnerabilities detected by \ossfuzz in real-world projects.

\input{figs/rq1Chart}

\textbf{Results.}
The pie charts in \autoref{fig: cmp_piechart} provide a comparative analysis of the patches for the three systems: \tool{}, \agentless and \vulmaster, respectively. 
All three charts categorize patches based on the aforementioned criteria, which are plausible (P), implausible (IP), compilation errors (CE), and No Patch (NP).
As can be seen from the left figure, 
\tool generated plausible patches for over half (52.6\%) of the vulnerabilities, showing the potential of \tool being used in real-world vulnerability remediation.
For the other 35\% of the vulnerabilities, \tool generated compilable but implausible patches.
The non-compilable patches account for a low percentage of the dataset (5.1\%), suggesting that the LLM can generated compilable patches to a large extent if it is given sufficient code context and type information.
For the remaining 6.8\% of the bugs, \tool did not generate a patch due to errors in the code search or other exceptions happened during the execution.

In comparison, the results from \agentless demonstrate that 49.3\% of the patches are implausible, which is higher than that from \tool.
In addition, 30.9\% of the patches are plausible (P), which is substantially lower than the plausible patch rate in \tool.
Furthermore, 11.6\% (68) of the patches result in compilation errors (CE), another clearly higher percentage compared to \tool. 
Lastly, for the remaining 8.2\% of bugs, \agentless fails to produce a patch.
In a nutshell, we observe that \tool achieves better efficacy than \agentless in all categories. 
To some extent, this is expected since \tool is targeted for security patching.

Unlike the agent systems that generated plausible patches for around 30-50\% of the vulnerabilities, \vulmaster could not generate plausible patches for most of the vulnerabilities.
The results from \vulmaster demonstrate that 63.6\% of the bugs do not have a compilable patch, the highest of all tools.
In addition, 30.4\% of the bugs do not have a patch that can be applied to the program.
Only 5.8\% of the subjects have an implausible patch. Finally, there is a single plausible patch, which is a function 
\begin{wrapfigure}[11]{r}{0.45\linewidth}
     \centering
     \includegraphics[width=1\linewidth]{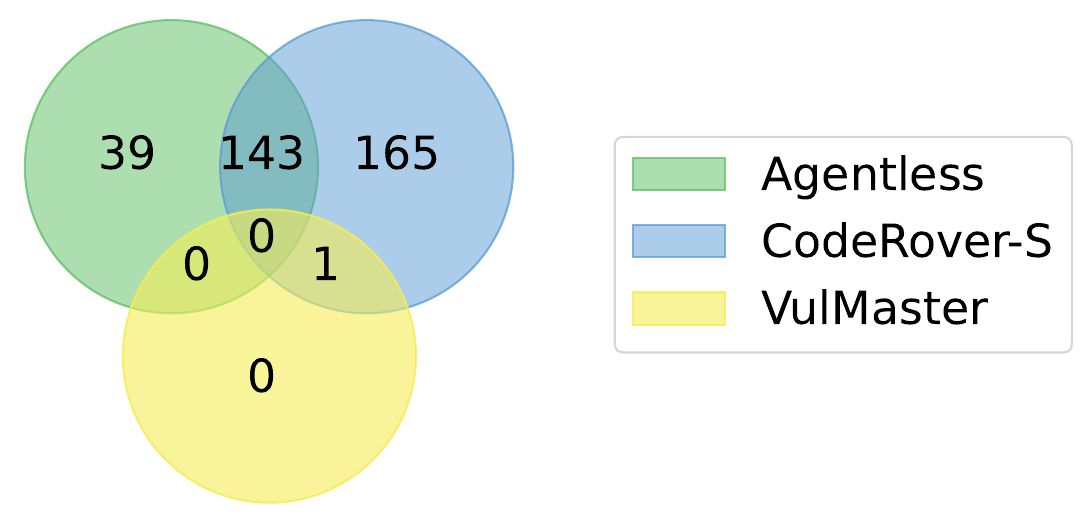}
     \caption{The Venn diagram for plausible patches from the three tools.}
     \label{fig:venn}
\end{wrapfigure}%
call removal. This suggests that \vulmaster is not capable of generating plausible patches
in realistic repair setups like \ossfuzz vulnerability remediation.

Furthermore, we examine to what extent plausible patches generated from the tools overlap. 
As seen in \autoref{fig:venn}, the overlap between \tool and \agentless is significant, with most of the patches (78.6\%) found by \agentless also present in that of \tool.
In summary, \tool demonstrates more consistency in producing plausible patches with fewer compilation errors, making it more reliable for generating useful security patches. 
\input{figs/lib-raw-patches}
\subsection{RQ2: Further Analysis of \tool Results}
\input{tables/cwe_fixes}
In this section, we provide further analysis of patches generated by \tool.
\paragraph{Fix rate across different categories}
For the vulnerabilities that \tool can find a plausible patch, we further examine how they are distributed across different bug types and years. 
\autoref{tab:fix_distribution_cwe} provides a breakdown of the plausible patch rate by the CWE types.
\tool performs better on buffer-related vulnerabilities, specifically CWE-121, CWE-122, and CWE-125, on which \tool achieved fix rates of 57\%, 69\% and 90\% respectively. These success rates suggest that \tool is adept at patching buffer-overflow and bounds-checking issues, which usually have well-defined patterns and remediation strategies.
In contrast, \tool struggles with other memory management-related vulnerabilities such as CWE-416 and CWE-457, where the fix rates drop to 35\% and 34\% respectively. These lower success rates may be attributed to the context-dependent nature of these vulnerabilities, as uninitialized variables and use-after-free may require a deeper semantic understanding of the program and the relationship between various variables, which could be a challenge for automated tools.

\input{tables/year_fix_cost_combined}

\autoref{tab:fix_distribution_year} presents the yearly distribution of \tool's plausible patch rates against the year of bug discovery, from 2016 to 2024. 
The analysis reveals that \tool's efficacy fluctuates within the range of 46\% to 63\% across vulnerabilities detected in different years (with 2016 being the outlier with only one vulnerability).
We further split the vulnerabilities detected in 2023 to ``Up-to-October'' and ``Post-October'', since the underlying \texttt{gpt-4o-2024-08-06} model used by \tool is only trained on data up to October 2023~\cite{gpt4o}.
Despite the underlying LLM has a knowledge cutoff date of October 2023, \tool still achieves similar efficacy in repairing vulnerabilities detected after October 2023 compared to previous years.
Notably, \tool achieves a higher efficacy of 63\% on vulnerabilities detected in 2024 compared to other years.
\tool's effectiveness on vulnerability remediation extends to vulnerabilities whose ground truth patches are certainly not included in the LLM's training data.

\paragraph{Time and cost}

We examine the average cost for an execution of \tool and the average time needed. 
The analysis in \autoref{tab:time_cost} shows the average and median time and cost for \tool's execution across different patch outcomes. For cases where a plausible patch was generated, the average time was significantly lower (23.23 minutes) compared to cases that resulted in an implausible (75.32 minutes) or non-compilable patch (44.58 minutes). 

In terms of cost, plausible patches are also the most economical, with an average cost of \$0.51 and a median cost of \$0.26. In contrast, implausible patches are more costly, averaging \$1.53, with a median of \$1.37, while non-compilable patches have an average cost of \$0.78 and a median of \$0.63. 
This indicates that \tool is more efficient in terms of cost and time when it successfully generates plausible patches, and this is likely due to fewer retry attempts, since the workflow stops once a plausible patch has been generated.

\begin{figure}[t]
    \centering
    \begin{subfigure}{0.47\textwidth} 
        \includegraphics[width=\textwidth]{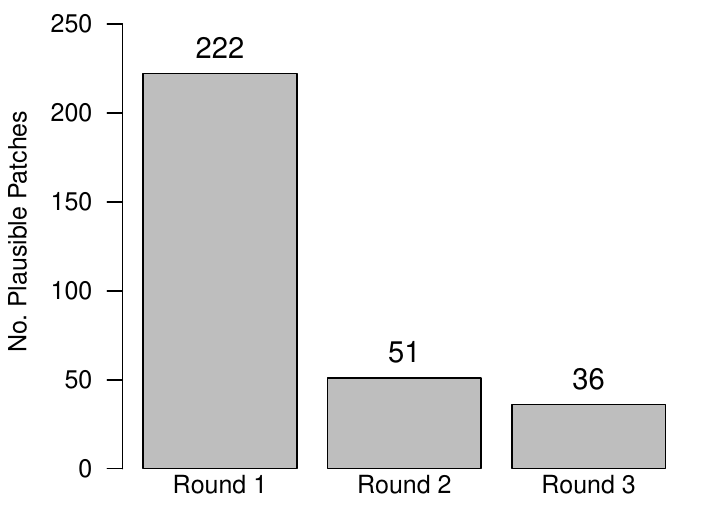}
        \caption{Number of bugs whose plausible patch is found in which round of the main loop.}
        \label{fig:plausible-patch-rounds-bar-chart}
    \end{subfigure}
    \hfill
    \begin{subfigure}{0.48\textwidth}  

        \includegraphics[width=\textwidth]{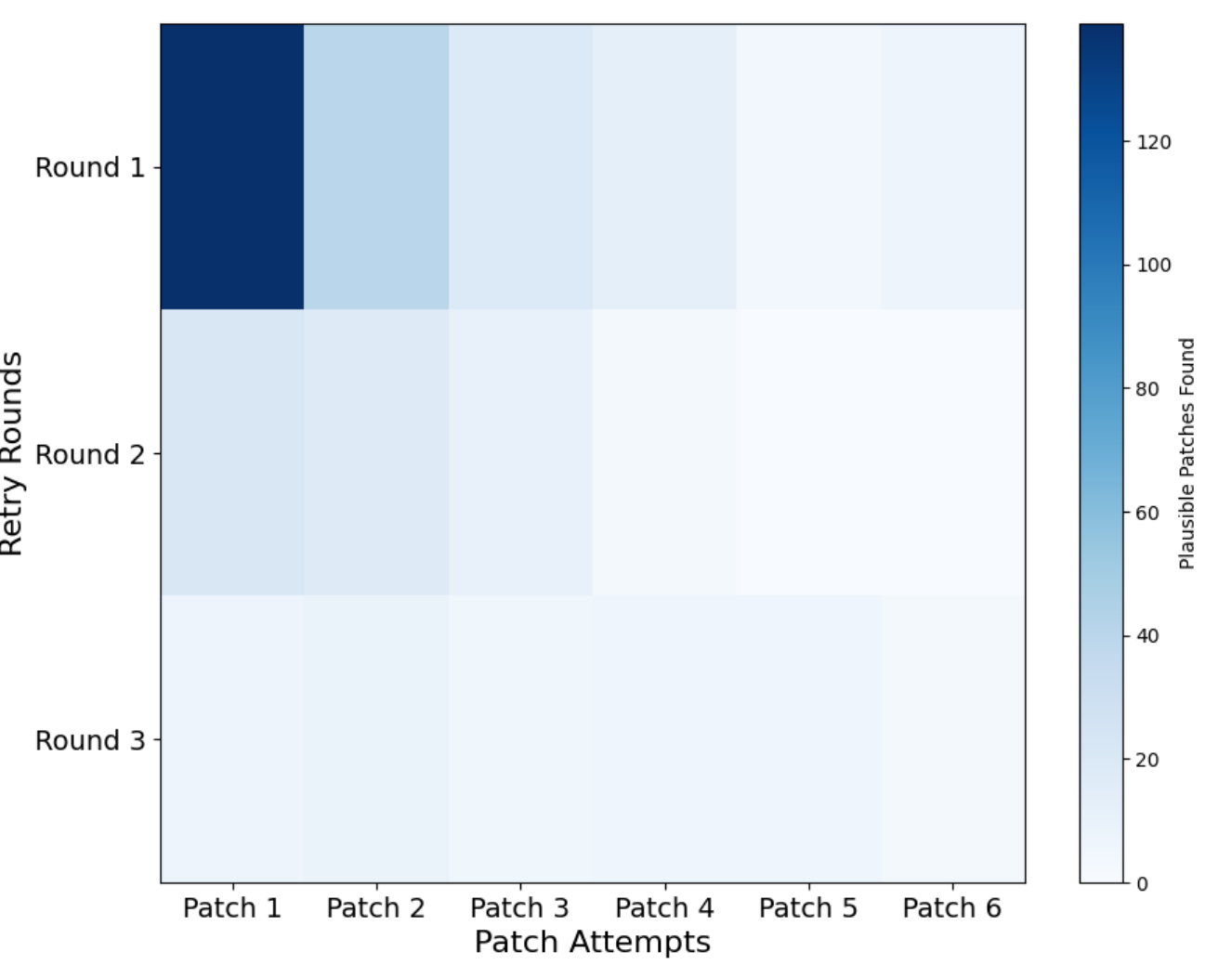}
        \caption{The heat-map of when the plausible patches are found, in more details.}
        \label{fig:plausible-patch-heatmap}
    \end{subfigure}
    \caption{Distribution of when the plausible patches are found.}
    \label{fig:plausible-patch-distribution}
   
\end{figure}

Furthermore, \autoref{fig:plausible-patch-distribution} provide insight into the distribution of when the plausible patch for each vulnerability was found.
In our experiments, \tool is configured to retry at most 6 times for patch generation and 3 times for the main loop (c.f. \autoref{fig:workflow_coderover-s}).
\autoref{fig:plausible-patch-rounds-bar-chart} shows that most of the plausible patches are generated in the first round of main loop, with diminishing results in later rounds. 
The heat-map in \autoref{fig:plausible-patch-heatmap} highlights that successful patches are primarily concentrated in the first few patch generation attempts of each round. 
While most plausible patches are identified during the first patching attempt in the initial round, a substantial number of plausible patches are still generated during the second attempt in both the first and second rounds. This highlights the effectiveness of the patch review process in refining the initial implausible patches.
In summary, both the plausible patch distribution and the time/cost analysis suggests that plausible patches are usually found at early stages in the \tool execution.
In potential real-world deployments, users can balance the efficacy and cost of \tool by establishing appropriate retry limits and execution timeouts.

\subsection{RQ3: Patch Validation Metric}

Other than the plausibility validation by executing tests, BLEU~\cite{papineni2002bleu} and CodeBLEU~\cite{ren2020codebleu} has been a popular metric for evaluating patch quality in AI-based vulnerability repair systems~\cite{zhou24vulmaster,islam2024code,wang2024navrepair}.
BLEU score computes n-gram similarity between two input sequences, and CodeBLEU additionally considers code syntax (via Abstract Syntax Tree, or AST) and semantics (via data-flow) of the two input code snippets.
Both of them were often used to estimate the quality of an auto-generated patch against the ground truth developer patch in previous studies, especially when executable tests are not available in the studied benchmark.
In this research question, we examine the extent to which high CodeBLEU scores can predict the actual quality of a patch, including factors like patch plausibility in vulnerability repair.
Therefore, our hypothesis formulation becomes: 
\begin{itemize}[leftmargin=*]
    \item \textbf{The null hypothesis} \textbf{$H_0$}: There is no significant association between CodeBLEU scores and the plausibility of code patches.
    \item \textbf{Alternative hypothesis} \textbf{$H_1$}: Higher CodeBLEU scores are significantly associated with plausibility of code patches.
\end{itemize}

To ensure the CodeBLEU score computation is meaningful, the generated solution and reference solution (i.e. developer patch) should modify similar locations in the program. 

Furthermore, since the CodeBLEU computation involves parsing the input sequence into AST, the input sequence should be a parsable code fragment.
To this end, we selected 124 vulnerabilities from the dataset used in RQ1, whose developer patch is single-hunk and is within a function.
We then pass in the developer patch location (which is a function) to \tool and instruct it to only generate patches at this location.
As a result, we obtain pairs of auto-generated patches and ground truth patches that only modify the same function.
For these 124 vulnerabilities under perfect localization conditions, \tool generated 111 compilable patches, of which 86 are plausible and 25 are implausible.
We compute CodeBLEU score for these 111 patches (i.e. 86 + 25) by extracting the patched functions after applying the auto-generated patch and the developer patch respectively.

\begin{figure}[ht]
     \centering
     \includegraphics[scale=0.4]{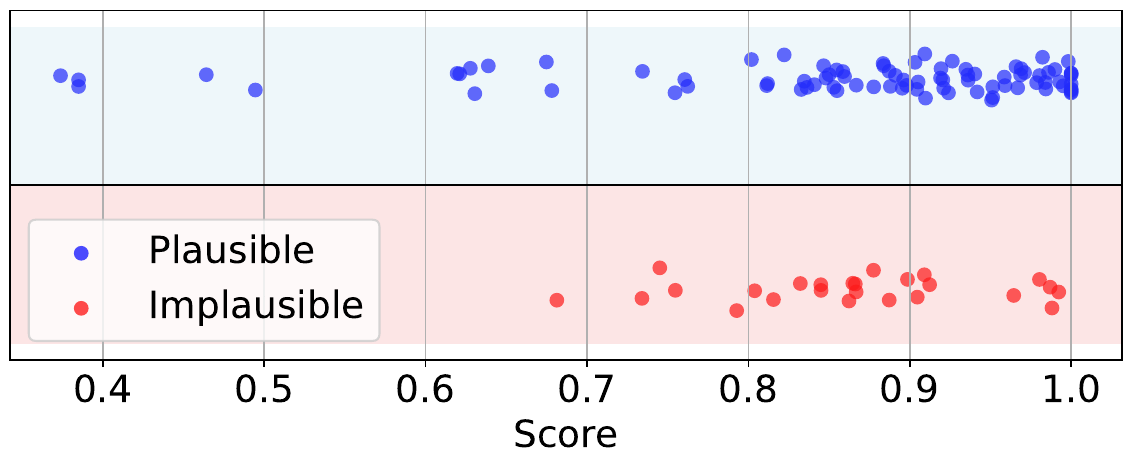}
     \caption{Point-Biserial Correlation coefficient is -0.008 and p-value is 0.94.}
     \label{fig:bleu_scatter}
\end{figure}

To test our hypothesis, we calculated the Point-Biserial correlation coefficient to assess the strength and direction of the monotonic relationship between the CodeBLEU scores and the categorical variable of plausibility. 
We choose this statistical method because the the plausibility of a patch is dichotomous in our definition.

\textbf{Results.} 
First of all, our results indicate no significant difference in the mean CodeBLEU scores between the implausible and plausible groups. 
The mean score for implausible patches is 0.878, and the mean score for plausible patches is 0.873. 
These scores suggest that CodeBLEU does not effectively distinguish between plausible and implausible patches in this context.
Then, we obtained a Point-Biserial correlation coefficient of -0.008 with a P-value of 0.94. 
This indicates that there is no statistically significant evidence of a linear relationship between CodeBLEU scores and the plausibility of patches. 
Specifically, we fail to reject the null hypothesis that no association exists between CodeBLEU and plausibility at the conventional significance level of 0.05. 
Furthermore, as shown in \autoref{fig:bleu_scatter}, the visualization of the data points reveals that even at lower CodeBLEU scores, there are many plausible patches, and at higher scores, there is a significant number of implausible patches.
These results indicate that, in the context of vulnerability repair, CodeBLEU does not reliably capture patch quality. Metrics like patch plausibility should be prioritized when executable tests are available.

%% file: figs/rq1Chart.tex
\begin{figure}[ht]
    \centering
    \definecolor{colorPlausible}{HTML}{4169E1} 
    \definecolor{colorImplausible}{HTML}{00CED1} 
    \definecolor{colorInapplicable}{HTML}{FFFF99} 
    \definecolor{colorCompilation}{HTML}{FFA07A} 

    \includegraphics[width=\textwidth]{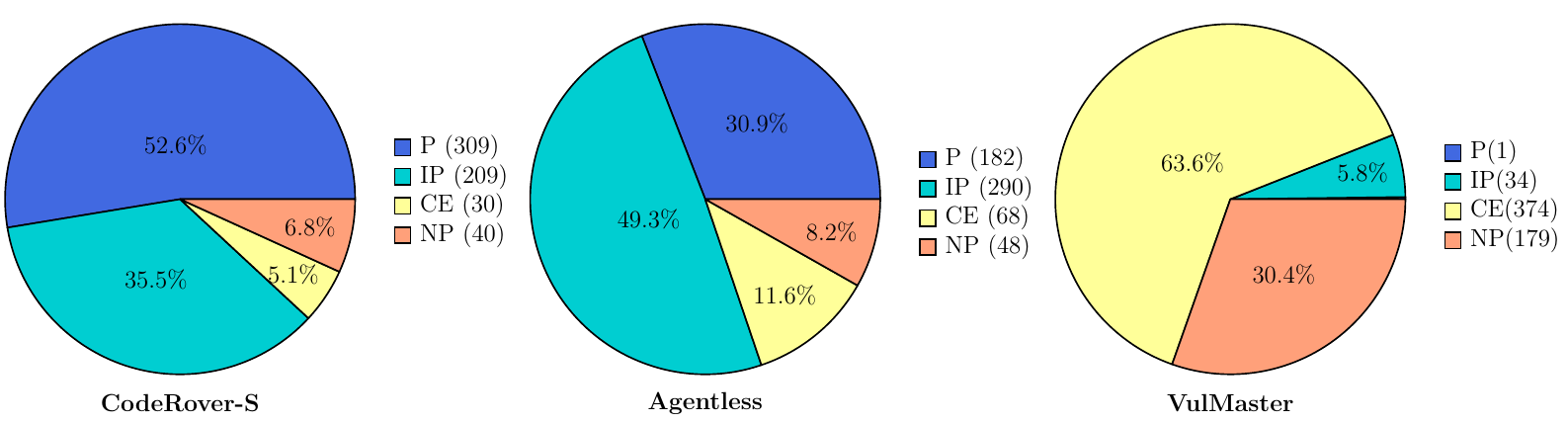}
    \caption{Comparison of results from \tool{}, \agentless and \vulmaster.}
    \label{fig: cmp_piechart}
\end{figure}

%% file: figs/lib-raw-patches.tex
\lstdefinelanguage{diff}{
  morecomment=[f][\color{blue}]{@@},     
  morecomment=[f][\color{red}]-,         
  morecomment=[f][\color{green}]+,       
  morecomment=[f][\color{magenta}]{---}, 
  morecomment=[f][\color{magenta}]{+++},
}

%% file: tables/cwe_fixes.tex
\begin{table}[h]
    \scriptsize
   
    \centering
    \begin{tabular}{lllll}
    \toprule
      CWE & \# Samples &  Name & \tool & \agentless \\
    \midrule
      CWE-121 &244 & Stack-based Buffer Overflow & 140 (57\%) & 77 (32\%) \\
      CWE-476 &105 & NULL Pointer Dereference & 42 (40\%) & 16 (15\%)  \\
      CWE-122 &67 & Heap-based Buffer Overflow & 46 (69\%) & 32 (48\%)  \\
      CWE-416 & 55 & Use After Free & 19 (35\%) & 15 (27\%)\\
      CWE-457 & 47& Use of Uninitialized Variable & 16 (34\%) & 15 (32\%) \\
      CWE-125 &20 & Out-of-bounds Read & 18 (90\%) &  13 (65\%) \\
      CWE-20  & 12 & Improper Input Validation & 8 (67\%) & 6 (50\%) \\
      CWE-626 & 12 & Null Byte Interaction Error & 6 (50\%) & 2 (17\%) \\
      CWE-120 & 11 & Buffer Copy without Checking Size of Input & 7 (64\%) & 4 (36\%) \\
      CWE-415 & 9 & Double Free & 4 (44\%) & 2 (22\%)\\
      CWE-590 & 5 & Free of Memory not on the Heap & 3 (67\%) & 0 (0\%)  \\
      CWE-770 & 1& Allocation of Resources  Without Limits or Throttling & 0 (0\%) & 0 (0\%)  \\ 
    \bottomrule
    \end{tabular}
     \caption{Analysis of plausible patches according to CWE Type.}
    \label{tab:fix_distribution_cwe}
\end{table}


%% file: tables/year_fix_cost_combined.tex
\begin{table}[t]
    \scriptsize
    \begin{minipage}{0.55\textwidth}
        \scriptsize
        \centering
            \begin{tabular}{lcccc}
            
            \toprule
            Year &\# Samples & \tool & \agentless &  \\
            \midrule
              2016 &1 & 1 (100\%) & 1 (100\%) &  \\
              2017 &30 & 15 (50\%) & 12 (40\%) &  \\
              2018 &59& 27 (46\%) & 17 (28\%) & \\
              2019 &57& 30 (53\%) & 10 (18\%) &  \\
              2020 &91& 48 (53\%) & 34 (37\%) &  \\
              2021 &138& 75 (54\%) & 45 (33\%) & \\
              2022 &80& 44 (55\%) & 21 (26\%) &  \\
              2023 (Up-to-October) &86& 43 (50\%) & 23 (27\%) &  \\
              2023 (Post-October) &22& 11 (50\%) & 8 (36\%) &  \\
              2024 &24 & 15 (63\%) & 11 (46\%) & \\
            \bottomrule
            \end{tabular}
            \caption{Analysis of plausible patches according to the year of bug detection.}
            \label{tab:fix_distribution_year}
    \end{minipage}%
    \hfill
    \begin{minipage}{0.4\textwidth}
        \scriptsize
        \centering
            \begin{tabular}{l|ll}
             
            \toprule
                & Time (mins)   & Cost (USD) \\
            \midrule
              No Patch &  8.25 (8.53) & 1.65 (1.63)  \\
              Compilation issue  & 44.58 (33.38) &  0.78 (0.63) \\
              Implausible          & 75.32 (51.88) &  1.53 (1.37) \\
              Plausible             & 23.23 (11.86) &  0.51 (0.26) \\
            \midrule
              All patches & 43.5 (24.83)  & 0.93 (0.69) \\
          \bottomrule
         \end{tabular}       
        \vspace{4pt}
       \caption{The mean and median for execution time and costs. Data in each cell are in the format of ``mean (median)''.}
        \label{tab:time_cost}
        
    \end{minipage}


\end{table}

%% file: related.tex
\section{Related Works}
\label{sec:related}
Automated Program Repair (APR) techniques \cite{cacm19} seek to heal faulty programs by automatically generating patches for developers. With the advent of AI-based coding,  attention on these techniques have further accentuated.
In this section, we describe AI-based repair and program analysis based repair, both of which are closely aligned with our work.

\subsection{Deep Learning and LLM-based Approach}
Various machine learning and deep learning based approaches have been proposed for program repair in the recent years.
The core idea is to model APR as neural machine translation (NMT) problem.
For instance, SequenceR~\cite{chen2018sequencer} uses neural networks to learn patterns from pairs of buggy and fixed code, enabling it to generate fixes end-to-end.
CoCoNuT~\cite{lutellier2020coconut} combines multiple neural network models for better representing the context of a bug.
Jiang et al.~\cite{jiang2021cure} further advance context-aware learning models by designing CURE. 
CURE handles the out-of-the-vocabulary issue and employs a code-aware search strategy (so that only valid identifiers are selected during inference).
Other than the general-purpose repair tools, several systems have been proposed for repairing security vulnerabilities.
VRepair~\cite{chen2021neural} is a vulnerability repair tool that is based on transfer learning. It is first trained on a large bug fix corpus, and then finetuned on a smaller vulnerability fix dataset. 
Instead of training a model and adapting it to vulnerability repair with transfer learning, VulRepair~\cite{fu2022vulrepair} finetunes a pre-trained CodeT5 model for repairing vulnerabilities.
VulMaster~\cite{zhou24vulmaster} is a recently proposed tool that integrates diverse information such as code structure and expert knowledge for vulnerability repair, and it is studied in our work.

In the era of LLMs, researchers have investigated using LLMs for program repair.
For example, ChatRepair~\cite{xia2023keep} proposed a conversational program repair workflow by interacting with ChatGPT.
Recently, various LLM agents have been proposed for autonomous program improvements and program repair.
\acr~\cite{zhang2024acr} uses a set of program structure aware search APIs for gathering relevant code context for the problem.
\sweagent~\cite{yang2024sweagent} designs an agent-computer interface which defines the possible actions taken by the agent for code viewing and editing.
\agentless~\cite{xia2024agentless} employs a fixed two-phase approach for localization and repair.
\agentless is studied in our work, and our tool \tool is built on top of \acr.
In this work, we present an experience of adapting autonomous agents for vulnerability repair on a large scale real-world dataset.

\subsection{Program Analysis based Approach}
A wide range of analysis-based approaches have been explored to repair software vulnerabilities~\cite{shariffdeen2021cpr,gao2021beyond,zhang2022vulfix,le2011genprog, mechtaev2016angelix}.
GenProg~\cite{le2011genprog} is one of the pioneering tools in APR. 
It uses genetic programming to evolve program variants and applies mutation and crossover operations to a program’s abstract syntax tree (AST) to generate candidate patches.
Moreover, SemFix by Nguyen et al.~\cite{nguyen2013semfix} combines symbolic execution with constraint solving to repair programs. 
It identifies program locations causing failures and generates patches by solving constraints that represent the desired program behavior.
 Gao et al.~\cite{gao2021beyond} explore the use of crash-free constraints to bootstrap the repair process. 
 The work leverages data-dependency localization to identify a precise location within the program's control flow graph, ensuring the vulnerability is effectively mitigated.
Zhang et al.~\cite{zhang2022vulfix} propose a counter-example guided inductive inference approach to define likely invariants at potential bug fix locations in a program.
Instead of relying on heavy symbolic analysis, VulnFix investigates using light-weight input/state generation to verify the candidate invariants, allowing good scalability.
These approaches provide strong correctness guarantees due to the rigorous program analysis foundations. 
However, they often suffer from scalability issues, especially with large codebases or complex software systems. Furthermore, the effort to move from one programming language to another is high in the analysis based approaches (as compared to LLM agent based approaches).

%% file: conclusion.tex
\section{Perspectives}

Automated program repair technologies typically involve search, semantic reasoning and machine learning to automatically rectify bugs and vulnerabilities. Typically such techniques are driven by a given test-suite as an indicator of correctness, and hence are difficult to apply for security vulnerabilities where only one test (the exploit) may be available. Recent emergence of Large Language Models (LLMs) have put the focus on fixing ``issues'' where natural language bug reports may be used to produce rectifying program modifications via LLM agents.  In this work, we have demonstrated the feasibility of using a repurposed LLM agent AutoCodeRover \cite{zhang2024acr} for rectifying security vulnerabilities found by continuous fuzzing taken from the widely used OSS-Fuzz infrastructure \cite{ossfuzz}. The main experience gained point us to the feasibility of using LLM agents as back-ends to fuzzers for zero-day patching of security vulnerabilities. Furthermore, such integration of fuzzers with LLM agents need to come with qualitative or quantitative measures of confidence in the fidelity of the patch. Recent work \cite{ruan2024specrover} have pointed to qualitative measures of confidence via explanations of patch generated from LLM agents. In future, one may also explore quantitative measures of confidence  in the patch by the agent's analysis of consistency of the different artifacts available to it, namely: tests, bug reports, method level specifications and patch candidates.